\documentclass[twocolumn]{aastex631}
\usepackage{booktabs}
\usepackage{float}
\usepackage{subcaption}
\usepackage{array}
\usepackage{float}
\usepackage{graphicx}	
\usepackage{amsmath}	
\usepackage{amssymb}
\usepackage{bm}		
\usepackage[T1]{fontenc}
\usepackage{ae,aecompl}
\usepackage{placeins}
\usepackage{newtxtext,newtxmath}

\newcommand{\tem}{t_{\rm emt}}
\newcommand{\tob}{t_{\rm obs}}
\newcommand{\nuob}{\nu_{\rm obs}}
\newcommand{\nuem}{\nu_{\rm emt}}
\newcommand{\betat}{\Tilde{\beta}}
\usepackage{xcolor}
\usepackage{hyperref}
\newcommand{\yisx}{\color{black}}
\newcommand{\yishuxu}{\color{black}}
\usepackage[normalem]{ulem} 

\begin{document}
\title{Long Pulse by Short Central Engine:\\Prompt emission from expanding dissipation rings in the jet front of gamma-ray bursts}
\correspondingauthor{Shu-Xu Yi} \email{sxyi@ihep.ac.cn} 
\author[0000-0003-1500-0002]{Shu-Xu Yi} 
\affiliation{Key Laboratory of Particle Astrophysics, Institute of High Energy Physics, Chinese Academy of Sciences, Beijing 100049, China} 
\author[0000-0002-4994-3365]{Emre Seyit Yorgancioglu} 
\affiliation{Key Laboratory of Particle Astrophysics, Institute of High Energy Physics, Chinese Academy of Sciences, Beijing 100049, China}
\author{S.-L. Xiong}
\affiliation{Key Laboratory of Particle Astrophysics, Institute of High Energy Physics, Chinese Academy of Sciences, Beijing 100049, People’s Republic of China}
\author{S.-N. Zhang}  
\affiliation{Key Laboratory of Particle Astrophysics, Institute of High Energy Physics, Chinese Academy of Sciences, Beijing 100049, People’s Republic of China}
\affiliation{University of Chinese Academy of Sciences, Chinese Academy of Sciences, Beijing 100049, China}



\begin{abstract}  
Recent observations have challenged the long-held opinion that the duration of gamma-ray burst (GRB) prompt emission is determined by the activity epochs of the central engine. Specifically, the observations of GRB 230307A have revealed a different scenario in which the duration of the prompt emission is predominantly governed by the energy dissipation process following a brief initial energy injection from the central engine. In this paper, we explore a mechanism where the energy injection from the central engine initially causes turbulence in a small region and radiates locally. This turbulence then propagates to more distant regions and radiates. Consequently, the emission regions form concentric rings that extend outward. Using an idealized toy model, we show that such a mechanism, initiated by a pulsed energy injection, can produce a prompt emission light curve resembling a single broad pulse exhibiting the typical softer-wider/softer-later feature. Under some parameters, the main characteristics of the GRB 230307A spectra and light curves can be reproduced by the toy model.  
\end{abstract}
\section{Introduction}
It is widely accepted that gamma-ray bursts (GRBs) have two distinct origins: the merger origin, linked to the mergers of compact binaries involving neutron stars—either neutron star-neutron star or neutron star-black hole mergers (Type I, \citealt{1974ApJ...192L.145L,1999ApJ...527L..39J,2006ApJ...648.1110B,zhang2006burst,2014ARA&A..52...43B}), and the collapsar origin, which is associated with the core collapse at the end of massive star evolution (Type II, \citealt{2003Natur.423..847H,2008ApJ...687.1201K,2012grb..book..169H}).
Observationally, GRBs can also be divided into two distinct categories based on their duration: short GRBs (sGRBs) and long GRBs (lGRBs). Until recently, it was commonly accepted that these two observational categories corresponded to the above mentioned two types of origins: sGRBs corresponded to mergers, while lGRBs corresponded to collapsars \citep{2006NatPh...2..116G,2008AIPC.1000...11B,2009A&A...496..585G,2013ApJ...764..179B,2012ApJ...759..107K}. The intuition behind such notion is that the duration of a GRB corresponds to the activity time of the central engine. In the case of collapsars, material from the massive star gradually falls into the newly formed compact object at the center, allowing the central engine to drive the lGRB for a longer period. In contrast, the merger process results in a shorter accretion time, which can only sustain the sGRB for a shorter duration. 

However, some observations in recent years have challenged this notion. For example, GRB 211211A and GRB 230307A, which have typical durations characteristic of lGRBs, were both detected with kilonova counterparts \citep{2022Natur.612..223R,2024Natur.626..737L}, which are considered key evidences for their origin from the mergers including neutron stars. Additionally, their other observational properties, such as the offset in their host galaxies and their positions in the $E_{\rm iso}$-$E_{\rm p}$ parameter space, also suggest that they should have a merger origin \citep{2022Natur.612..232Y,sun2023magnetar,2024arXiv240702376W}. But how do mergers drive a longer-duration GRB like these bursts\footnote{See a recent comprehensive discussion over this topic by \cite{2025JHEAp..45..325Z}.}, if one does not appeal to novel merger companions such as white dwarfs \citep{2022Natur.612..232Y}, or magnetars \citep{2022ApJ...939L..25Z}?. 

Recently, \cite{yi2023evidence} pointed out that the duration of GRB 230307A's prompt emission is independent of the activity timescale of its central engine. Instead, the entire prompt emission corresponds to a dissipation process associated with an impulsive energy injection from the central engine. Now, if the duration of GRBs can be independent of the activity time of the central engine itself, a question arises: how can we explain the dichotomy of GRBs' durations?
From our perspective, the duration of a GRB depends on the maximum value of two timescales: the dissipation timescale of energy injection from a single epoch of central engine activity and the duration of the central engine's multi-epoch activity. To address the previously mentioned question, 
we need to answer the following: what determines the dissipation timescale of an impulsive energy injection? 

For GRB 230307A, the study by \cite{yi2023evidence} demonstrates that its prompt emission is characterized by a broad pulse, which is composed of numerous correlated smaller pulses. These individual small pulses correspond to localised radiation processes in different causally connected regions. They point out that if the entire radiation region is dominated by magnetic energy, then Alfvén waves in the magnetic field can naturally act as a mediator for this causal connection. Therefore, they demonstrate that the Internal-Collision-Induced Magnetic Reconnection \citep[ICMART,][]{zhang2010internal} radiation mechanism might best explain the prompt emission of GRB 230307A. In this framework, the duration of the entire broad pulse should be determined by the timescale over which these causal connections propagate throughout the entire radiation region. In the ICMART model, the typical radius of the prompt emission is $R\sim10^{15-16}$ cm. Thus, the size of the radiation zone is $\theta_{\rm J} R$ and the co-moving frame timescale for Alfvén waves to propagate throughout this region is:

\begin{equation}
  \tau^\prime = \frac{\theta_{\rm J}R}{c}\,.
\end{equation}

Substituting $\theta_{\rm J}\sim10^{-2}$, we find $\tau^\prime\sim10^3$. Therefore, from the observer's perspective, the time scale of the GRB can be $\tau\sim\tau^\prime/\Gamma\sim10^2$ s. This demonstrates that the spreading timescale of turbulence that triggers local radiation throughout the emission region naturally corresponds to the aforementioned dissipation timescale. 

In this paper, we will carefully calculate the observational phenomena induced by this physical process. As a starting point, we will consider a simplified scenario, where the emission region of the prompt emission is a geometrically thin shell in a relativistic jet. When this shell reaches a certain radius, a small localised region on the shell begins to radiate. It is important to note that, unlike in the usual treatments of high-latitude effect arguments, the entire shell does not shine simultaneously;  instead, only a single point on the shell radiates initially. This luminous point then extinguishes shortly after the initial pulse, which has generated a disturbance that propagates causally outward from that point within the shell. In the regions reached by this disturbance, new localised pulses of emission emerge.


In the next section, we will rigorously define our model mathematically, and then we will calculate in detail how the specific flux observed by an observer varies with time in this scenario. In the next section, we assume that the energy spectra of the localized radiation processes adhere to specific physical mechanisms. We then use these spectra to simulate the time-resolved spectrum and energy-resolved light curves observed by the observer. In the following section, we will discuss how our model relates to real observations. Specifically, we discuss what of the observed characteristics of GRB 230307A can be reproduced by our model, and what can not. {\yisx The codes for generating the figures are linked with a symble `</>' in the caption of the corresponding figure.}  

\section{The model}\label{sec:intro}
\begin{figure}
  \centering
  \includegraphics[width=8 cm]{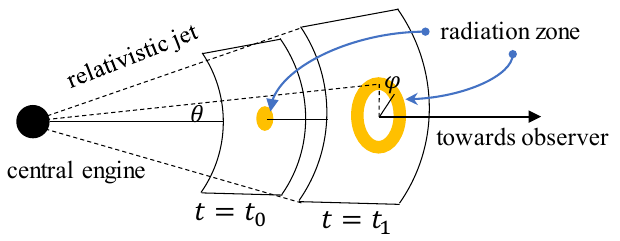}
  \caption{The illustration of the model. {\yisx In this scenario, the emission happens in a propagating ring region within a thin spherical shell. The shell represents the radius of the jet, where the dissipation of the magnetic field energy happens. }}
  \label{fig:model}
\end{figure}

We are considering such an emitting source in the relativistic jet: the emission is radiated from a thin sphere, whose radius is $R$ and expands outwards. The locations on the sphere are described by spherical coordinates $\theta$ and $\varphi$. Instead of emitting as a whole at the same instant, we consider the following case: the source sphere is illuminated as a pulse at $\theta=0$ initially, and the illuminated region propagates outwards to larger $\theta$ at instant $t$. The $\theta$ of the illuminated ring area is a function of $t$. See figure \ref{fig:model} for an illustration of the model. Denote the propagating velocity of the "illuminating wave" along the sphere in the source co-moving frame as $v^\prime=\Tilde{\beta}c$. Therefore, in the co-moving frame, one has:
\begin{equation}
    \Tilde{\beta}c=\frac{d(R\theta)}{dt^\prime}=\theta\frac{dR}{dt^\prime}+R\frac{d\theta}{dt^\prime}.
\end{equation}
Since the sphere is also expanding, we have $dR/dt=\beta c$ in the rest frame. Using the relation $dt=\Gamma dt^\prime$, we have:
\begin{equation}
    \betat c=\theta\Gamma\beta c+\Gamma\beta ct\frac{d\theta}{dt}.
    \label{eq:diff}
\end{equation}
Solving the differential equation with the initial condition that $\theta=0$ at $t=t_0$, we have:
\begin{equation}
\theta=\frac{\betat}{\beta\Gamma}(1-t_0/t),\qquad \text{for}\,t>t_0.
\label{eq:thetat}
\end{equation}
We can therefore write the specific emission coefficient $j^\prime_{\nu^\prime}$ as:
\begin{equation}
j^\prime_{\nu^\prime}=\epsilon f(\nu^\prime)\delta(r-R)\delta(t-t_\theta)\cos\theta,
\label{eq:source}
\end{equation}
where the Dirac delta function $\delta(r-R)$ corresponds to the assumption that the source is a spherical shell with infinitesimal thickness, and $\delta(t-t_\theta)$ corresponding to our assumption that the ring at $\theta$ is only illuminated at the instant when the illuminating wave propagated to $\theta$ for a pulse. $f(\nu^\prime)$ is the normalised emitting spectrum in the co-moving frame, where 
$$\int f(\nu^\prime)d\nu^\prime=1,$$ and $\epsilon$ is the factor proportional to emitting power. {\yisx The $\cos\theta$ term is to account for the plate-like geometry of the emitting volume element, whose the project area is proportional to $\cos\theta$.}

Knowing the specific emission coefficient in the source rest frame, we can calculate the spectrum flux to be observed with (see Appendix):
\begin{equation}
F_{\nuob}(\tob)=\frac{\Gamma}{D^2_{\rm{L}}}\int\mathcal{D}^3 j^\prime_{\nuem^\prime}(\mathbf{r},\tem,\hat{\Omega})dV.
\label{eq:fv2p}
\end{equation}
In the above equation, we explicitly denote the difference between the photon receiving time $\tob$ and its emitting time $\tem$. The relation between $\tob$ and $\tem$ is that: 
\begin{equation}
    \tob=(1+z)(1-\beta\cos\theta_t)(\tem-{\tem}_0).
    \label{eq:tt}
\end{equation}
Taking equation (\ref{eq:source}) into the above equation, we obtain:
\begin{eqnarray} \nonumber
&&F_{\nuob}(\tob)\\ \nonumber
&=&\frac{2\pi\Gamma}{D^2_{\rm{L}}}\int\mathcal{D}^3\epsilon f(\nuem^\prime)r^2\delta(r-R)\delta(t-t_\theta)\cos\theta\sin\theta dr d\theta\\ \nonumber
&=&\frac{2\pi\Gamma}{D^2_{\rm{L}}}\int\mathcal{D}^3\epsilon f(\nuem^\prime)R^2\delta(t-t_\theta)\cos\theta\sin\theta d\theta\\ \nonumber
&=&\frac{2\pi\Gamma}{D^2_{\rm{L}}}\int\mathcal{D}^3\epsilon f(\nuem^\prime)R^2\delta(t-t_\theta)\cos\theta\sin\theta\frac{d\theta}{dt_{\theta}}dt_{\theta}\\ 
&=& \frac{2\pi\Gamma}{D^2_{\rm{L}}}\mathcal{D}(t)^3\epsilon f(\nuem^\prime)R(t)^2\sin\theta_t\cos\theta_t\frac{d\theta_t}{dt}.
\end{eqnarray}
In the above equation, $\theta_t$ is the $\theta$ as a function of $t$ as defined in equation (\ref{eq:thetat}). 

We further assume that the emission power $\epsilon$ is proportional to the local energy density of the ejecta $U$, which is inversely proportional to the $R^2$. Hence, $\epsilon=\epsilon_0(R/R_0)^{-2}$, and bringing this into the above equation, we have:
\begin{equation}
    F_{\nuob}(\tob)=\frac{2\pi\Gamma\epsilon_0R^2_0}{D^2_{\rm{L}}}\mathcal{D}(t)^3 f(\nuem^\prime)\sin\theta_t\cos\theta_t\frac{d\theta_t}{dt}.
    \label{eq:Fv}
\end{equation}

\begin{figure}[htbp] 
  \centering 
  \begin{subfigure}[t]{0.45\textwidth} 
    \centering \includegraphics[width=\textwidth]{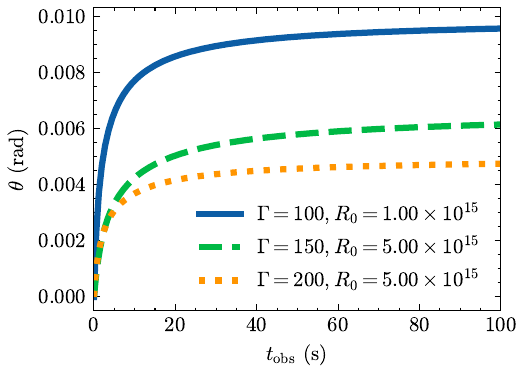} 
  \end{subfigure}
  \hspace{0.05\textwidth} 
  \begin{subfigure}[t]{0.45\textwidth} 
    \centering 
    \includegraphics[width=\textwidth]{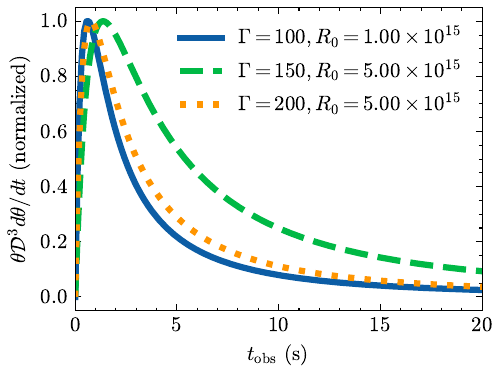}
  \end{subfigure}
  \caption{$\theta$ and the product $\theta\mathcal{D}^3d\theta/dt$ as function of $\tob$ (upper and lower panels respectively) {\yisx under different choices of parameters. \href{https://code.ihep.ac.cn/sxyi/long_pulse_by_short_central_engine/-/blob/main/figures/check-parameters.ipynb}{</>}}}
  \label{fig:twin}
\end{figure}

From equations (\ref{eq:thetat},\ref{eq:tt}), we can calculate the corresponding $\tob$ from each given $\tem$. In the upper panel of figure \ref{fig:twin}, we show $\theta$ as a function of $\tob$. 

As we can see from the upper panel of figure \ref{fig:twin}, $\theta$ initially increases rapidly, and then approximates to a maximum angle. From equation \ref{eq:thetat} we know that this asymptotic angle is $\theta\rightarrow\frac{\betat}{\beta\Gamma}$. It can be understood that as the jet front expands, the angular horizon size is $\frac{\betat}{\beta\Gamma}$, beyond which, the perturbation at one site cannot propagate to other sites in the jet front. 

In equation (\ref{eq:Fv}) we can see the specific flux $F_\nu$ depends on $\sin\theta\cos\theta\propto\theta$, which contributes to a fast increase at early times. On the other hand, $F_\nu$ depends also on $\mathcal{D}^3$ and $d\theta/dt$, which decrease with $\tob$. The interplay of these three factors will give an overall fast-increase-slow-decrease shape, as we shown in the lower panel of figure \ref{fig:twin}. 
\section{Emission Spectrum}
Now we will consider the intrinsic emission spectrum in the co-moving frame $f(\nu^\prime_{\rm{emt}})$. We consider a phenomenological spectrum with piecewise powerlaws:
\newcommand{\nup}{{\nu^\prime}}
\begin{equation}
  f(\nup) \propto 
  \left\{ \begin{array}{ll}
                \nup^{1/3} & \nup \le \nu_{\rm turn}(R) \\
                \nup^{\Tilde{\alpha}(R)} & \nu_{\rm turn}(R) \le \nup \le \nu_{\rm break}(R) \\
                \nup^{-p/2} & \nup \ge \nu_{\rm break}(R)
          \end{array} \right.
          \label{eq:emit_spectrum}
\end{equation}
where the turning and breaking frequencies, as well as the index of the middle frequency range are all functions of the radius of the shell (we avoid using $\alpha$ to distinguish it from the commonly used photon index of the Band spectrum). This spectrum shape aligns with the fast-cooling weak self-absorption regime \citep{zhang2018physics}, whereas we introduce more flexibility by allowing the index of the middle frequency to vary. $p$ is the power index of the injected electron energy distribution. As the shell expands, we expect the characteristic frequencies to decrease as the local magnetic field decays:
\begin{equation}
  \begin{array}{cc}
  \nu_{\rm turn} = & \nu_{\rm turn, 0}R^{-a}\\

  \nu_{\rm break} = & \nu_{\rm break,0}R^{-b}.
  \end{array}
  \label{eq:11}
\end{equation}

In observations of GRB 230307A, the middle segment of the $\nu f_{\nu}$ spectrum was found to flatten with time \citep{sun2023magnetar}. We therefore introduce a phenomenological evolution of $\Tilde{\alpha}(R)$ to mimic this behavior:

\begin{equation}
  \Tilde{\alpha}(R)=\frac{R_0}{2R}-1.
  \label{eq:12}
\end{equation}
With this formulism, $\Tilde{\alpha}(R)$ is initially  $-1/2$ as in the standard fast-cooling regime, but gradually becomes $-1$ as indicated from the observation of GRB 230307A. 

\begin{figure}
  \centering
  \includegraphics[width=8 cm]{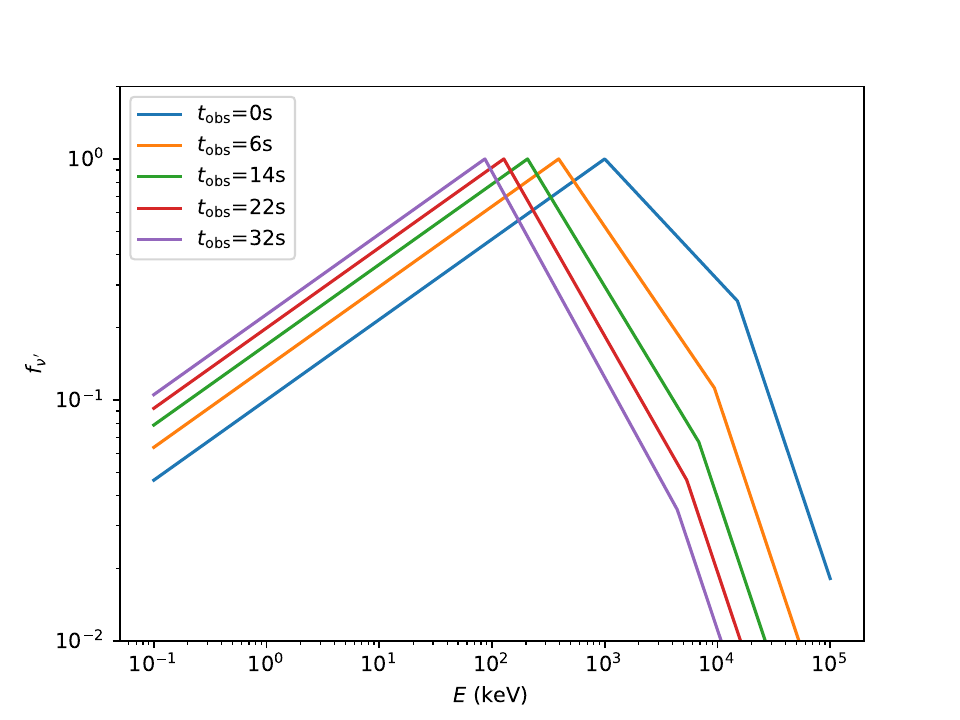}
  \caption{The emission spectrum in the co-moving frame {\yisx in different time in the observers frame, as described in equations \ref{eq:emit_spectrum}. The parameters used when generating the spectral are described in equations (\ref{eq:11},\ref{eq:12}). \href{https://code.ihep.ac.cn/sxyi/long_pulse_by_short_central_engine/-/blob/main/figures/spectrums.ipynb}{</>}}}
  \label{fig:fv}
\end{figure}

We plot the emitting spectra $f_{\nup}$ in different instants in figure \ref{fig:fv}. When plotting figure \ref{fig:fv}, we use $p=2.8$ (in accordance with magnetic reconnection \citep{uhm2014fast, sironi2014relativistic, guo2014formation}, $h\nu_{\rm turn,0}=1 \text{keV}$ and $h\nu_{\rm break,0}=15\,\text{keV}$, $a=2$, $b=1$. {\yishuxu As shown in Figure \ref{fig:fv}, the emitting spectrum constructed using the aforementioned formulation is represented by a three-segment power law, with both turnover frequencies decreasing over time. This may represent the physics that the cooling energy and the minimum injection energy of electrons at the synchrotron radiation site decrease over time due to the dilution of the magnetic field strength within the expanding shell.}

\section{Simulated time-evolving spectrum and energy resolved light curves}

In this section, we integrate the intrinsic emission spectrum into equation (\ref{eq:Fv}) to obtain the energy resolved lightcurve and the time-evolving spectrum. The time-evolving spectrum is defined in a time window from $t_{\rm start}$ to $t_{\rm end}$:
\begin{equation}
F(\nuob)=\int^{t_{\rm end}}_{t_{\rm start}}F_{\nuob}(\tob) d\tob/(t_{\rm end}-t_{\rm start}).  
\end{equation}
and the energy resolved time light curve is defined in an energy range from $h\nu_{\rm start}$ to $h\nu_{\rm end}$:
\begin{equation}
  F(\tob)=\int^{\nu_{\rm end}}_{\nu_{\rm start}}d\nu.
\end{equation}

\begin{figure*}[b]
  \centering
  \includegraphics[width=\textwidth]{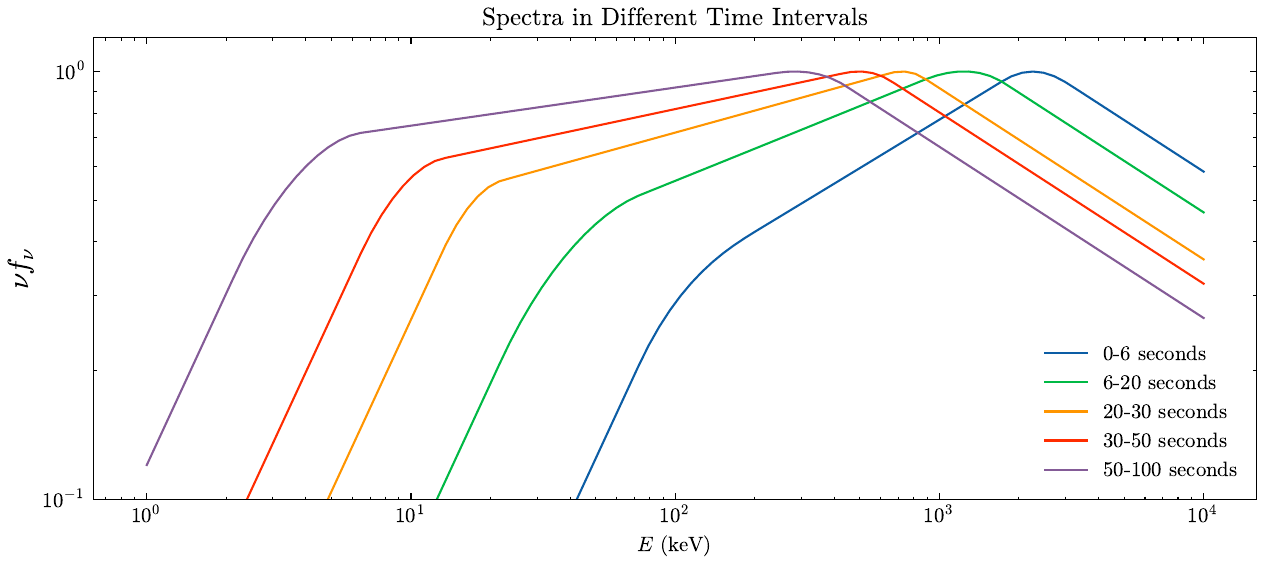}
  \caption{\yisx Averaged spectrum over different observer's time intervals. The corresponding time intervals are listed in the legend of the figure. \href{https://code.ihep.ac.cn/sxyi/long_pulse_by_short_central_engine/-/blob/main/figures/spectrums.ipynb}{</>}}
  \label{fig:fvt}
\end{figure*}

In the following calculation, we apply the following parameters: $\Gamma=100$, $R_0=5\times10^{15}$ cm, $\betat=0.99$ and $z=0.06$ (the same as GRB 230307A); the spectrum parameters are $p=2.8$, $\nu_{\rm turn, 0}=1$ keV, $\nu_{\rm cut,0}=15$\,keV, $a=2$ and $b=1$. The resulting time-evolving spectrum is plotted in figure \ref{fig:fvt}. {\yishuxu The observed spectrum as a function of time follows a three-segment power law. The decrease in the peak energy corresponds to decrease of the $\nu_{\rm break}$ in the emitting spectrum, which is to represent the softening of the spectrum over time in observation. The flattening of the middle segment of the spectrum over time is phenomenologically reproduced by Equation \ref{eq:12}.}

\begin{figure*}[t] 
  \centering 
  \begin{subfigure}[t]{0.48\textwidth} 
    \centering \includegraphics[width=8.4 cm]{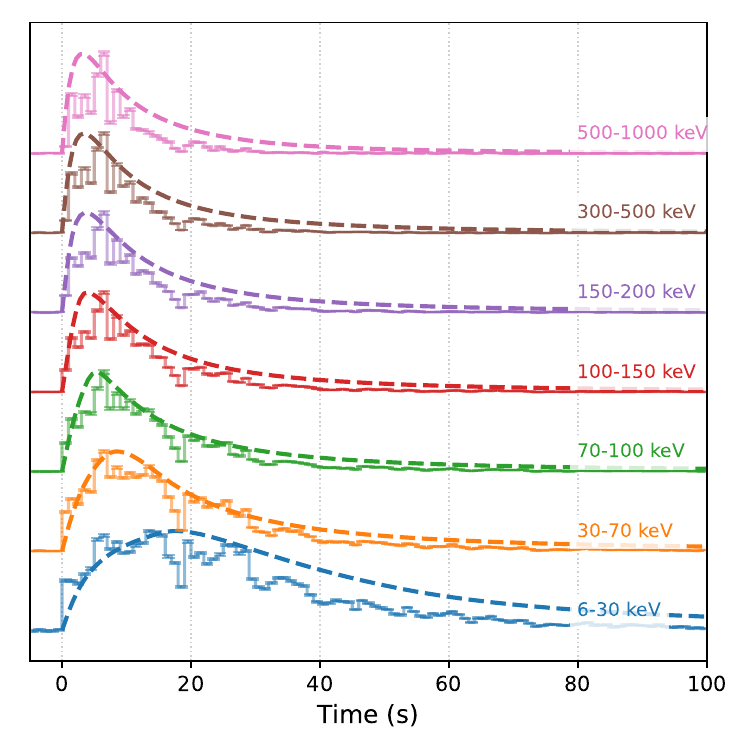} 
  \end{subfigure}
  \begin{subfigure}[t]{0.5\textwidth} 
    \centering 
    \includegraphics[width=9.2 cm]{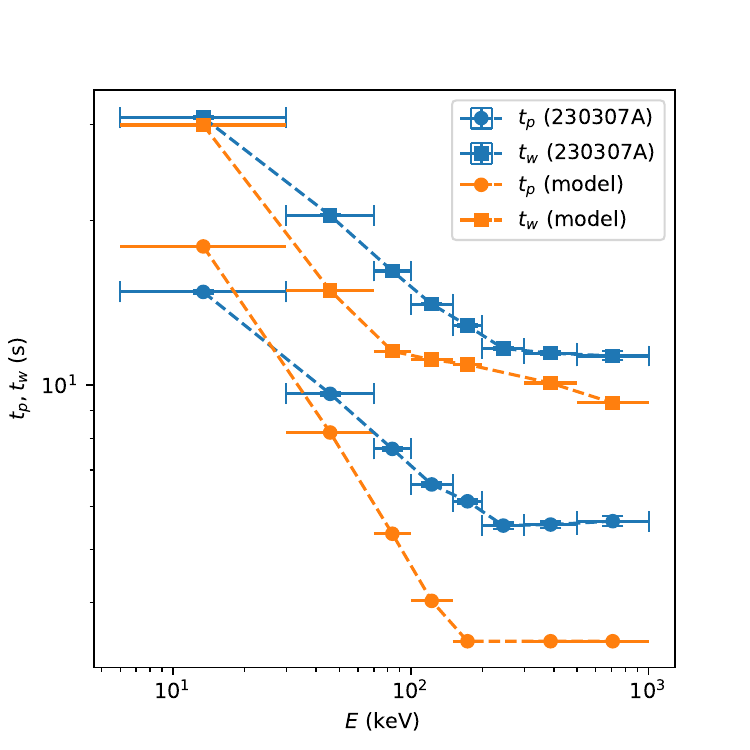}
  \end{subfigure}
  \caption{\textbf{Left Panel:} {\yisx The simulated light curves in different energy bands (curves) vs. the observed ones in GRB 230307A (histograms);} \textbf{Right Panel:} The peaking time and the width of the light curves as function of the energy. {\yisx observed ones from GRB 230307A are plotted along (blue) for comparison. The horizontal error bars denote the energy ranges over which the light curves are calculated, and the markers locate in the geometrical averaged energy of the corresponding band. The dashed lines link the markers in order to guide the eyes for the trends. \href{https://code.ihep.ac.cn/sxyi/long_pulse_by_short_central_engine/-/blob/main/figures/light_curves.ipynb}{</>}}}
  \label{fig:lcs}
\end{figure*}

The energy resolved light curves are plotted in the left panel of figure \ref{fig:lcs}. {\yisx We overlap the corresponding data from GRB 230307A for comparison to the model. The data are taken from \cite{yi2023evidence}.}
In the right panel of figure \ref{fig:lcs}, we plot the energy dependence of the peaking time $t_p$ and the width $t_w$ (defined as the width at the $1/e$ of the peak flux) of the light curves. {\yisx The observed values are also plotted along in the same figure.} {\yishuxu We can see in figure \ref{fig:lcs} that, some key observational features of GRB 230307A can be reproduced, such as the broad FRED pulse profile, the softer-wider/softer-later behaviour, the power law dependency of energy on $t_p,t_w$  and the saturation of this dependency above a certain energy. Those above mentioned features can be reproduced over a range of model parameters. However, some discrepancies remain. For instance, the strict self-similarity observed in GRB 230307A can not naturally reproduced in the simulation: the slopes and the saturation energy of the $E-t_p$ and $E-t_w$ relationships are identical in the observation of GRB 230307A, but this is difficult to fully reproduce in the simulation. Additionally, the slopes of the energy dependence are shallower in GRB 230307A than those from the simulations.}



\section{Conclusion and Discussion}
With this toy model, we demonstrate that the propagation of turbulence within the dissipation region can naturally extend the emission process, producing a long pulse from a brief energy injection by the central engine.  Although simplified in several aspects (we will elaborate later), this model can still reproduce many key features of GRB 230307A:
\begin{enumerate}
  \item A broad FRED (Fast-Rise-Exponential-Decay) pulse prompt emission light curve;
  \item Self-similar light curves in multiple energy bands, with softer-wider and softer-later behavior;
  \item The softer-wider and softer-later behavior saturates at higher energies;
  \item The spectrum softens with time, with a decreasing $E_{\rm p}$.  
\end{enumerate}
Most of the reproduced features are robust under a range of model parameters, while the self-similarity among the lightcurves in different energy bands are not in general guaranteed. By self-similarity, \cite{yi2023evidence} means the shape of lightcurves in different energy bands can be rescaled in time to be identical with each other. In our model, such a behavior can only be reproduced with small ranges $\nu_{\rm turn,0}$, $\nu_{\rm break,0}$, $a$, $b$. Therefore, the observed self-similarity of GRB 230307A can still not be naturally explained, until one can find the physical reason for the emission spectrum quantitatively.    

In order to substitute this toy model with a physical model, one need to replace several simplified components of this model: 
\begin{enumerate}
  \item The Dirac delta time-dependence $\delta(t-t_{\theta})$ of the emissivity should be treated in a more physically realistic manner, taking into account the cooling process of electrons;
  \item The phenomenological emitting spectrum should be replaced by a physical spectrum, where we need to solve the Fokker-Planck equation \citep{chandrasekhar1943stochastic, petrosian2012stochastic} in each ring;
  \item In real observations of GRB 230307A, the long broad pulse is composed of many short pulses, which can be explained as local random magnetic reconnections. In our toy model, we assume a uniform emissivity throughout the jet front, and therefore the lightcurve is smooth. If we replace the emissivity with localized Dirac delta functions in $\theta$ and $\phi$, the short pulses can be reproduced too. 
\end{enumerate}
We will leave the aforementioned tasks to future work. 

All of the above calculations and discussions are based on two crucial presumptions: first, that the initial illumination occurs exactly at \( \theta = 0 \); and second, that only a single point is initially illuminated. We expect that if the initial point is slightly off-axis (\( \theta \ne 0 \)), it would distort the Fast Rise Exponential Decay (FRED) shape of the lightcurve and alter the monotonic softer-wider or softer-later behavior. A rigorous treatment of this case would involve a more complicated geometry but would still fit within the framework of this toy model. On the other hand, considering multiple initially illuminated points complicates the scenario further, as turbulence would propagate differently in regions where the magnetic field energy has already been dissipated by earlier arriving rings. In general, we anticipate that such a scenario would result in faster overall dissipation, thereby producing a shorter prompt emission in the GRB with a more complex lightcurve.


\FloatBarrier
\section*{Acknowledgments}
This work is supported by the fund from the Chinese Academy of Sciences (grant Nos. E32983U810). We also acknowledge funding support from the National Natural Science Foundation of China (NSFC) under grant No. 12333007.
\newpage
\appendix
\section{From source specific emission coefficient to flux spectrum at detector}
Denote the specific emission coefficient as $j_{\nuem}(\mathbf{r},\tem,\hat{\Omega})$, which is defined as the energy $dE$ emitted at $\tem$ per time interval $d\tem$, from per source volume element $dV$ at location $\mathbf{r}$, at around frequency $\nuem$ per frequency interval $d\nuem$, towards a certain direction $\hat{\Omega}$ per solid angle $d\Omega$:
\begin{equation}
\centering
j_{\nuem}(\mathbf{r},\tem,\hat{\Omega})\equiv\frac{dE}{d\tem dVd\Omega d\nuob}.
\end{equation}

On the other hand, the flux spectrum $F_{\nuob}(\tob)$ is defined as the energy received per unit time interval $d\tob$ per area $d\sigma$ per frequency interval $d\nu$:
\begin{equation}
F_{\nuob}(\tob)=\frac{dE}{d\tob d\nuob d\sigma}
\end{equation}
For a detector at a large distance $D_{\rm L}$ to the source, the solid angle opened by the detectors projected area $d\sigma$ is: $d\Omega=d\sigma/D^2_{\rm L}$.  Therefore, we have:
\begin{equation}
F_{\nuob}(\tob)=\frac{1}{D^2_{\rm{L}}}\int j_{\nuem}(\mathbf{r},\tem,\hat{\Omega}) \frac{d\tem}{d\tob}\frac{d\nuem}{d\nuob}dV.
\end{equation}
When considering a source moving at relativistic velocity $\beta c$, we have $d\tob=(1+z)(1-\beta\cos\theta)d\tem$, and $d\nuob=d\nuem/(1+z)$, where $\theta$ is the angle between the jet velocity and the light of sight. Note that the emission time $\tem$ and observed $\tob$ are measured with local clocks which are all in the rest frame respect to the observer (neglecting the Hubble expansion). Therefore, the factor $(1-\beta\cos\theta)$ which differentiates $d\tem$ and $d\tob$ originates purely from geometrical effects, that the location of the source has changed between $\tem$ and $\tem+d\tem$, so that the propagating times of photons from these two instants are different. Therefore, the above equation becomes:
\begin{equation}
F_{\nuob}(\tob)=\frac{1}{D^2_{\rm{L}}}\int(1-\beta\cos\theta)^{-1}j_{\nuem}(\mathbf{r},\tem,\hat{\Omega})dV.
\label{eq:fv}
\end{equation}

The intrinsic properties of the emission mechanism of the source determine the specific emission coefficient in the source comoving frame. We denote this with:
\begin{equation}
j^\prime_{\nuem^\prime}\equiv\frac{dE^\prime}{d\tem^\prime dV^\prime d\Omega^\prime d\nuem^\prime};    
\end{equation}
Since we know that the four dimensional volume element $dtdV$ and the ratio $dE/d\nu$ are Lorentz invariants, and the solid angle element $d\Omega=\mathcal{D}^{-2}d\Omega^\prime$, we therefore have: 
\begin{equation}
    j_{\nuem}=\mathcal{D}^2j^\prime_{\nuem^\prime},
    \label{jv}
\end{equation}
where $\mathcal{D}$ is the Doppler factor $\mathcal{D}=\frac{1}{\Gamma(1-\beta\cos\theta)}$, and $\Gamma=\frac{1}{\sqrt{1-\beta^2}}$.
Taking equation (\ref{jv}) into equation (\ref{eq:fv}), we have:
\begin{equation}
F_{\nuob}(\tob)=\frac{\Gamma}{D^2_{\rm{L}}}\int\mathcal{D}^3j^\prime_{\nuem^\prime}(\mathbf{r},\tem,\hat{\Omega})dV.
\label{eq:fv2}
\end{equation}
Note that we are still writing $j^\prime_{\nuem^\prime}$ as function of time, coordinates and direction vector in the rest frame. 
\bibliography{ref}{}
\end{document}